\documentclass[aps,pre,reprint]{revtex4-1}

\newcommand{\twoversions}[2]{#2}
\usepackage{amsmath}
\usepackage{graphicx}
\usepackage[usenames, dvipsnames]{xcolor}
\usepackage[normalem]{ulem} 
\newcommand{\added}[1]{{\twoversions{\color{blue}{#1}}{#1}}}

\newcommand{\deleted}[1]{{\twoversions{\color{blue}\sout{#1}}{}}}

\twoversions{
\def\keyFont{\fontsize{8}{11}\helveticabold }
\def\firstAuthorLast{Mello} 
\def\Authors{Bernardo A. Mello}

\begin{document}
\onecolumn
\firstpage{1}

\title[One-way pedestrian traffic]{One-way pedestrian traffic is a means of reducing personal encounters in epidemics} 

\author[\firstAuthorLast ]{\Authors} 
\address{} 
\correspondance{} 

\extraAuth{}

\maketitle
}{
\begin{document}
\title{One-way pedestrian traffic is a means of reducing personal encounters in epidemics}
\author{Bernardo A. Mello}
\affiliation{Physics Institute\\University of Brasilia\\Brasilia, DF, Brazil, 70919-970}
\date{\today}
}

\begin{abstract}
Minimizing social contact is an important tool to reduce the spread of diseases, but harms people's well-being. This and other, more compelling reasons, urge people to walk outside periodically. The present simulation explores how organizing the traffic of pedestrians affects the number of walking or running people passing by each other. By applying certain rules this number can be significantly reduced, thus reducing the contribution of person-to-person contagious to the basic reproductive number, $R_0$.  One example is the traffic of pedestrians on sidewalks. Another is the use of walking or running tracks in parks. It is demonstrated here that the number of people crossing each other can be drastically reduced if one-way traffic is enforced and runners are separated from walkers.
\twoversions{\tiny
 \keyFont{ \section{Keywords:} Epidemics, pedestrians, jogging, urban mobility, basic reprodutive number}}{} 
\end{abstract}

\twoversions{}{\maketitle}

Contagious epidemics,  as the Covid-19 pandemic, often demands limiting physical interactions among people in order to reduce the contagious rate. Governmental measures to reduce physical contact range from the closing of public facilities and schools to restrictions on mobility, lockdowns, quarantines, and curfews. These extreme measures though necessary, should be used as last resources, due to their economic and personal negative impacts. Of great help in these situations are the physical and psychological benefits of physical exercises, walking included~\cite{lee2008importance, roe2011restorative}. On the other hand, physical contact and proximity should be avoided to reduce the spread of pathogens such as Covid-19~\cite{sohrabi2020world}.

Measures that reduce the physical interaction with minimal disruption in the daily activities, as the ones proposed here, must be adopted whenever possible. For example, if sidewalk and crosswalks at intersections are made one-way, with walking only allowed on the right-hand ones (the street must be at pedestrian's left), the major inconvenience would be one more block of walking for pedestrians when reaching their destinations. 

Pedestrian behavior depends on internal and external aspects such as urban environment, contingent individual situation, and crowd behavior~\cite{zacharias2001pedestrian, seitz2016cognitive, von2014humans}, which must all be considered when planning traffic interventions. Sophisticated methods have been proposed to study pedestrian motion~\cite{von2015dynamic, seitz2012natural, feliciani2016improved} and used, for example, to studies the efficiency of pedestrian mobility~\cite{davidich2013predicting}. Mostly, the studies on urban mobility have concentrated on efficiency, well being, safety, and other relevant aspects of daily life~\cite{goodwin1999transformation, middleton2010sense,st2014happy}. 

The present work addresses a completely different goal, which is only justifiable in abnormal circumstances, such as epidemics: minimizing the crossings among pedestrians. The conceptual problem is much simpler, since the interpersonal effect on mobility is not relevant due to the low density of people, justifying the use of a more `pedestrian' mathematical model.

\section{Simulating the crossing of walkers}
\begin{figure}
\begin{center}
\mbox{\includegraphics[scale=0.35]{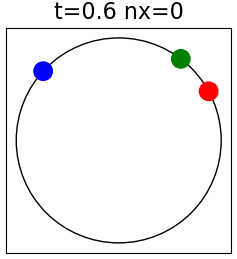}
\includegraphics[scale=0.35]{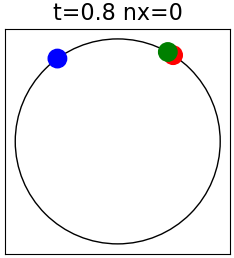}}\\
\mbox{\includegraphics[scale=0.35]{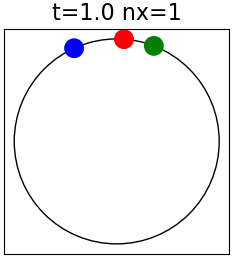}
\includegraphics[scale=0.35]{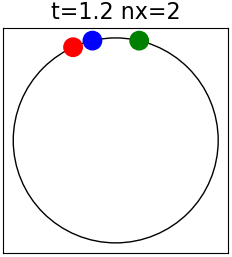}}\\
\mbox{\includegraphics[scale=0.35]{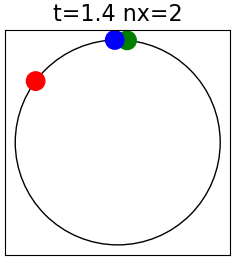}
\includegraphics[scale=0.35]{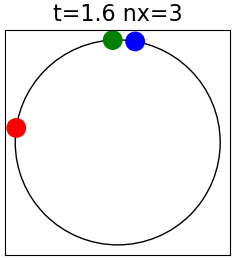}}\\
\end{center}
\caption{Example of crossings between walkers on a circular track. The blue walker moves in the clockwise direction and the green and the red walkers move in the counter-clockwise direction, the red walker moving faster than the green. The time and the number of crossings up to that instant is shown above each figure.} \label{fig:dynamics}
\end{figure}

Figure~\ref{fig:dynamics} illustrates the movement of three walkers on a circular track. The first crossing occurs between $t=0.8$ and $1.0$, involving two walkers moving in the same direction, the faster red  overpassing the slower green. The second crossing occurs between $t=1.0$ and $1.2$, between the blue and the red walkers that move in opposite directions. The crossing of walkers moving in the same direction is due to different speeds, and its frequency is reduced if the walkers walk at a similar rate, regardless of fast or slow. On the other hand, the crossing frequency of walkers moving in opposite directions is proportional to the average of their absolute speeds, regardless of the difference in their absolute values.

To simulate the movement of a crowd around a track, we assign a different and constant speed to each person. Measures of walking speed in several conditions presented in \cite{chandra2013speed} were used to adopt the mean value of 1.4~m/s with the standard deviation of 0.25~m/s for the walking speed. From \cite{smyth2018fast}, the running speed is assumed to be twice these values, i.e., an average of 2.8~m/s and standard deviation of 0.5~m/s. Random speeds with normal distribution are assigned to each walking or running person, with the corresponding average and standard deviation.

Two measures are proposed to evaluate the number of crossings per person. One is the number of crossings per minute, that is suitable to evaluate the crossings of a person who goes out for a given amount of time, for example, for jogging. The other is the number of crossings when a person walks along 100~m, typically, the length of one block. This is suitable in the analysis of a person who goes out to reach a certain place.

Each simulation presented here covers 1 hour of 500 people strolling in a 5\,000~m track. The results are the average number of crossings per person at each minute or at each block (100~m). These two quantities are proportional to the average density of people on the track, in this case, 1 person at every 10~m, and do not depend on the simulation time or the track length. Proportionality may be used to determine these quantities for other densities.

\section{Methods}
A circular track was simulated and a random generator with an uniform distribution over the track length was used to define the initial position. One random generator with normal distribution $\mu=1.4$~m/s and $\sigma=0.25$~m/s was used to assign a constant speed to each walker. Similarly for the runners,  with $\mu=2.8$~m/s and $\sigma=0.5$~m/s. When minimum or maximum speed were imposed, the speed of individuals under or above these limits was redefined as equal to the boundary values. When the track was bidirectional, a binary random generator was used to reverse the speed of roughly half of the population.

\begin{figure}
\begin{center}
\includegraphics[width=7.2cm]{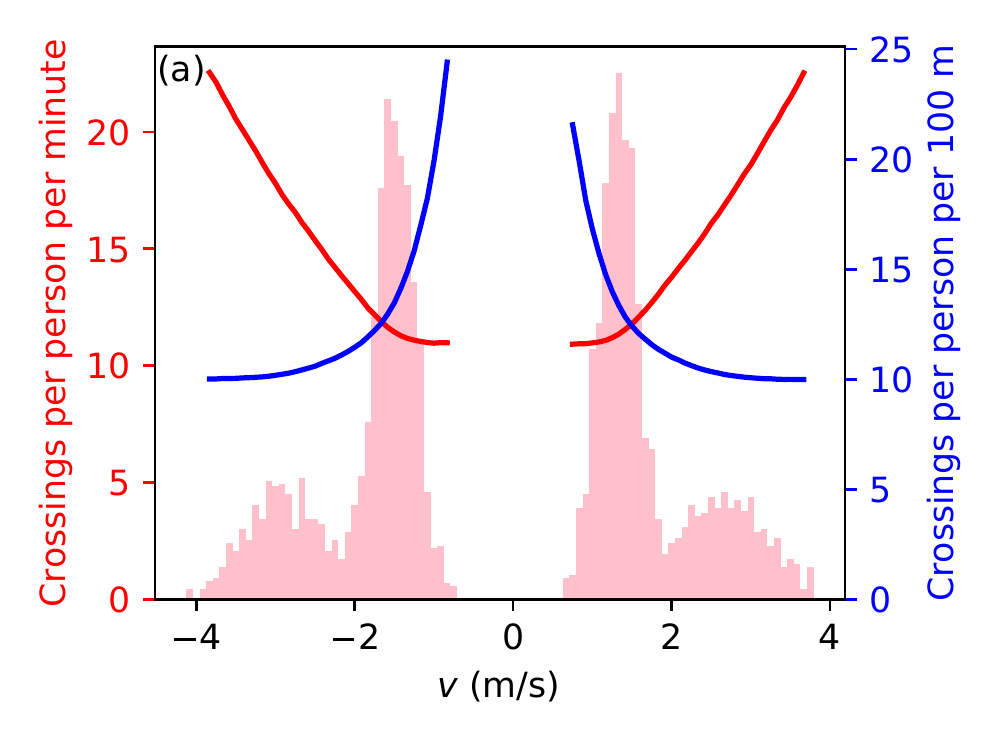}\\
\includegraphics[width=7.2cm]{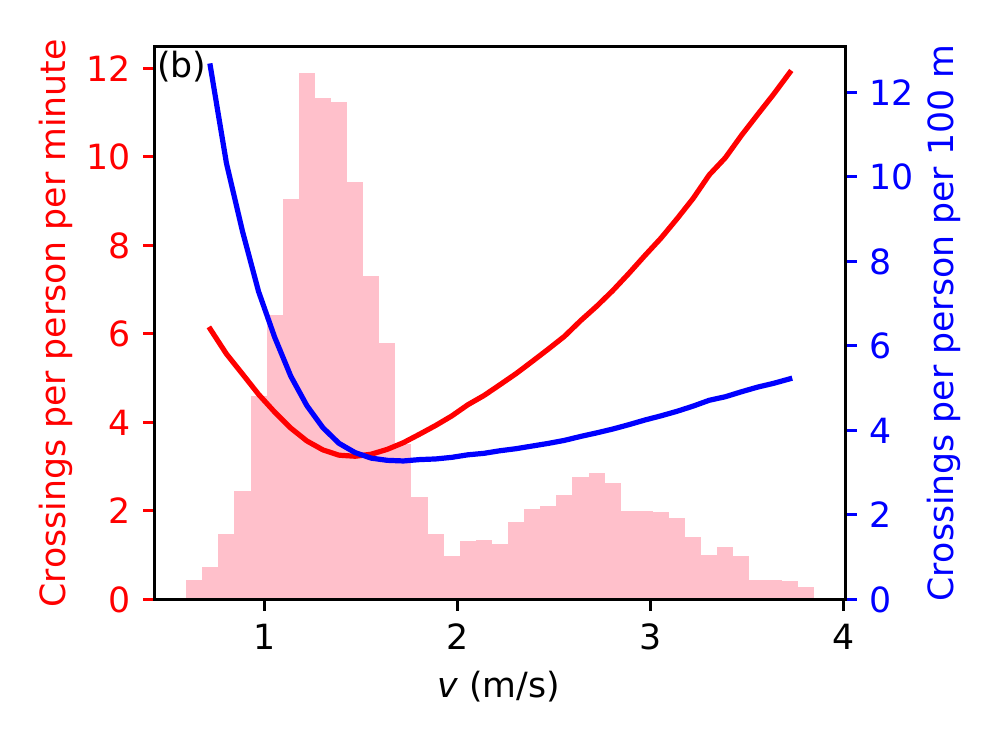}\\
\includegraphics[width=7.2cm]{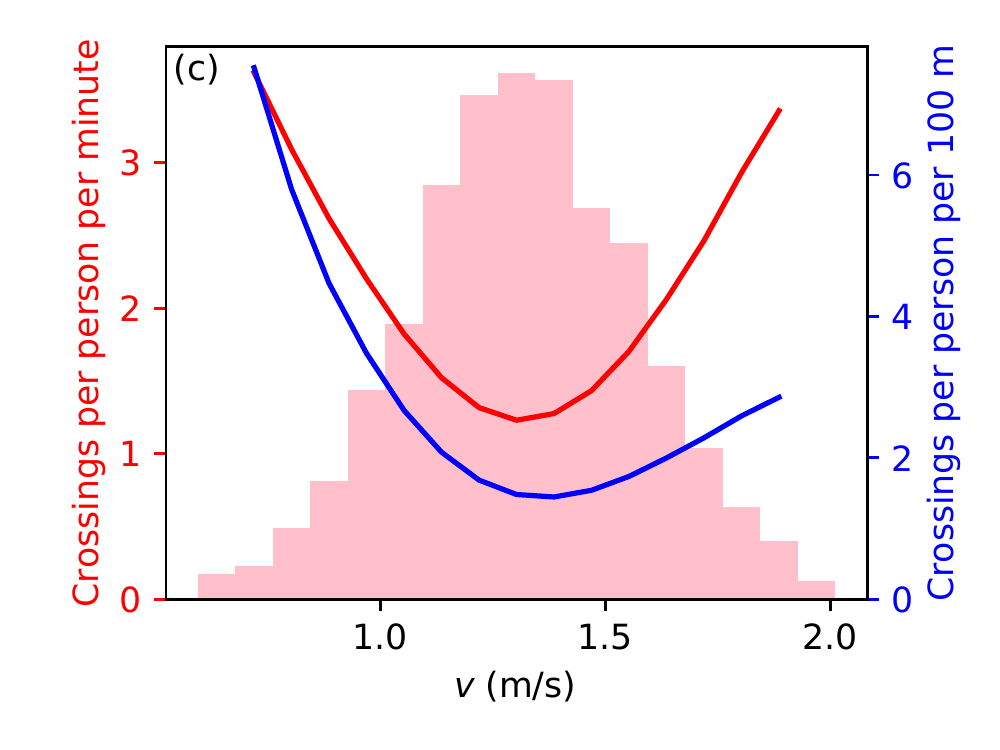}\\
\end{center}
\caption{a) Simulation of a 5\,000~m track shared by 500 people moving on both directions, 70~\% of walkers and 30~\% of runners. The distribution of speeds among the strollers is shown as pink bars in arbitrary units. The red curve is the average number of crossings per minute for people moving at a certain speed. The blue curve is the average number of crossings per 100~m for people moving at a certain speed. b) Same as (a), with one-way enforced. c) Same as (a) with one-way enforced and running forbidden.}
\label{fig:cross_v}
\end{figure}

The time evolution was performed by Euler integration with constant $\Delta t$. To avoid the evaluation of $N^2$ pairs of pedestrians when checking the occurrence of crossings, the track was divided into segments of length $(v_\text{max}-v_\text{min})\Delta t$, and only individuals in the same or in neighbor segments where checked. From the number of crossings of each individual, $\times_i$, the mean number of crossings per minute and per 100~m were, respectively, calculated as
\begin{equation}
\times_\text{minute} = 60 \, \frac{\left< \times_i \right>}{T_\text{simulation}} 
\end{equation}
and
\begin{equation}
\times_\text{100~m} = 100 \, \Bigl< \frac{\times_i}{L_i} \Bigr>,
\end{equation}
where $T_\text{simulation}$ is the total simulation time, in seconds, and $L_i$ is the distance traveled by the individual $i$, in meters. The averages were taken over the whole population when calculating the data for Table~\ref{table} or over groups within intervals of speeds when preparing the plots in Fig.~\ref{fig:cross_v}.

To reduce the statistical noise, the presented data were obtained from simulations with 4\,000 individuals on a 40\,000~m track, which leads to the same mean values as 500 individuals on a 5\,000 m track.

\section{Walking rules and the number of crossings}
Our first simulation is a track with equal numbers of people moving in both directions, 70~\% of them walking and 30~\% running. These percentages were arbitrarily chosen but the results are qualitatively equivalent for other values. The distribution of their speeds is shown in figure~\ref{fig:cross_v}a, together with the number of crossings per minute and per 100~m. It can be seen that the fast runners run by several people per minute, but have a minimal number of encounters along 100~m since they cover that distance quickly. A slow walker, on the other hand, exhibits the opposite behavior, for reverse reasons. In both cases, most of the encounters are between people moving in opposite directions.


To reduce the number of crossings observed in Fig.~\ref{fig:cross_v}a, we enforce unidirectional movement, illustrated by Fig.~\ref{fig:cross_v}b. The maximum and the minimum values of the number of crossings per minute and per 100~m are reduced to less than half of their previous values. By comparing the bidirectional and the unidirectional columns of line 3 in Table~\ref{table}, it can be seen that the number of crossings per minute and per 100~m are both reduced by 65~\%, to around one third of their values in bidirectional traffic.

The encounters between people moving unidirectionally  are due to their heterogeneous speed, which can be made more homogeneous by separating runners and walkers. When compared to Fig.~\ref{fig:cross_v}b, Fig.~\ref{fig:cross_v}c shows a substantial reduction in the crossing rates at any  speed for walkers only, moving one-way.  By comparing the columns of unidirectional encounters of lines 3 and 5 of Table~\ref{table}, we find a reduction to less than half in both cases. Therefore, forbidding running among walkers also reduces the number of crossings.

\begin{table}
\begin{center}
\begin{tabular}{|c|c|c|c|c|c|c|c|}
\hline
& &  & Fraction & \multicolumn{2}{|c|}{$\times_\text{minute}$} & \multicolumn{2}{|c|}{$\times_\text{100 m}$} \\
\cline{5-8}
& $v_\text{min}$ & $v_\text{max}$ & of runners & Bidir. & Unidir. &  Bidir. & Unidir. \\
\hline
1&--&--& 100 \%&18.48&3.36&11.23&2.12\\ 
2&--&--& 50 \%&15.4&5.46&13.26&4.95\\ 
3&--&--& 30 \%&13.2&4.65&12.94&4.49\\ 
4&--&--& 10 \%&10.7&2.91&12.04&3.13\\ 
5&--&--& 0 \%&9.25&1.74&11.25&2.18\\ 
6&--&1.65& 0 \%&9.03&1.47&11.09&1.9\\ 
7&1.15&--& 0 \%&9.25&1.46&10.97&1.72\\ 
8&1.15&1.65& 0 \%&9.02&1.23&10.83&1.5\\ 
\hline
\end{tabular}
\end{center}
\caption{The average number of crossings per minute and per 100~m for 500 pedestrians in a 5\,000 m track subjected to different conditions. The $v_{min}$ and $v_\text{max}$ are the minimum and the maximum acceptable speeds for people on the track. The third column is the fraction of runners on the crowd.} 
\label{table}
\end{table}

Line 1 of Table~\ref{table} shows the result of the simulation for a track where only runners are allowed. Comparison with the walkers on line 5 of the same table shows that the number of crossings per 100~m is slightly lower for runners, but the number of crossings per minute is around twice bigger for runners. The density of walkers on a track of runners must be half of the density on a track of walkers to produce the same number of crossings per minute.


Further reduction in the crossing rates can be achieved by imposing minimum and maximum walking speeds, for example, to no more and no less than one standard deviation of the average speed. Reductions of 30~\% are obtained, as can be seen by comparing the unidirectional crossings of lines 5 and 8 in Table~\ref{table}. The modest improvement and the practical difficulties of imposing such measures suggest that such measures should not be applied. 

\begin{figure}
\begin{center}
\includegraphics[width=9cm]{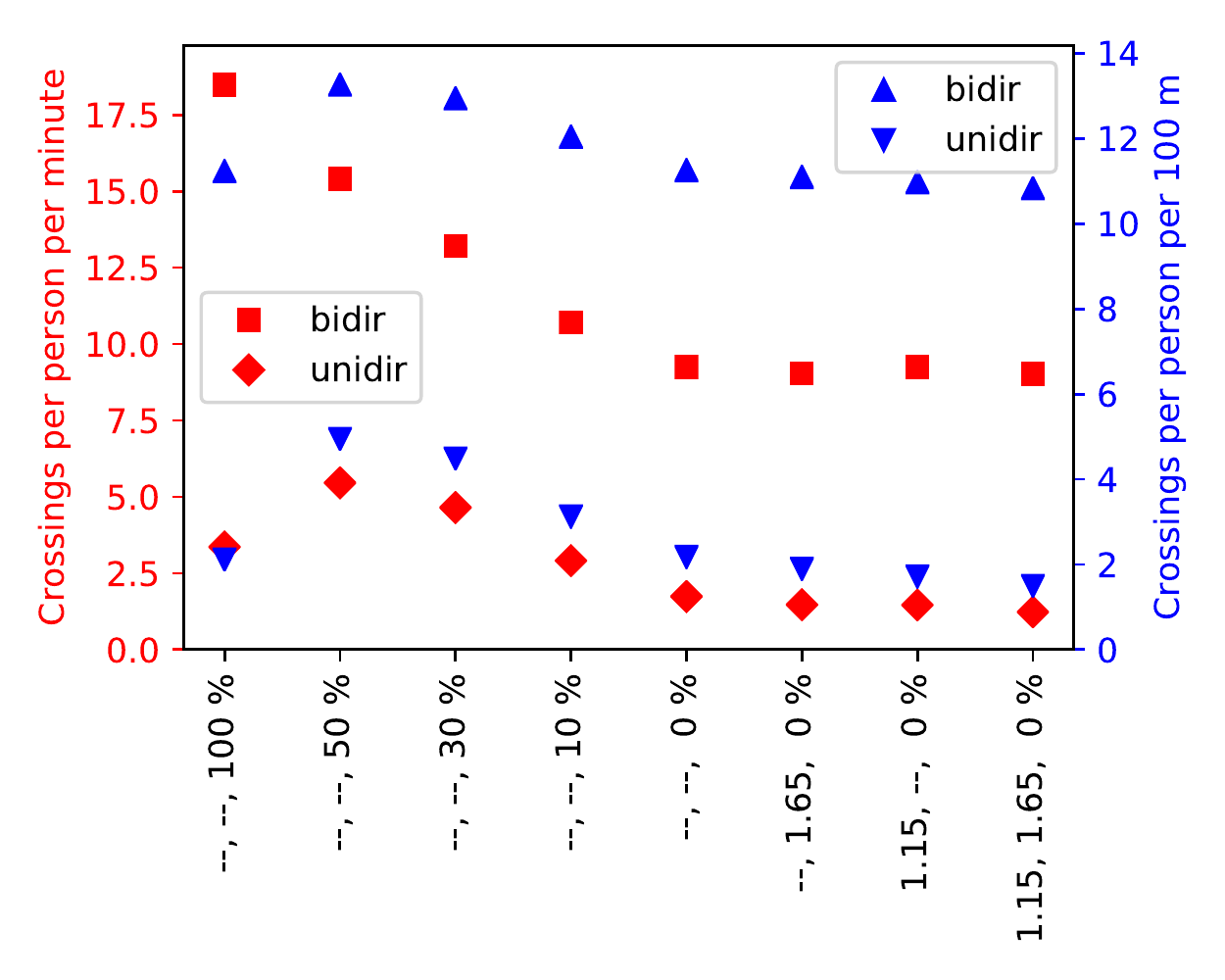}
\end{center}
\caption{Plotting of the data presented in Table~\ref{table}.}
\label{fig:crossings}
\end{figure}

Results of simulations in other conditions were performed as presented in Table~\ref{table} and illustrated in Figure~\ref{fig:crossings}. It provides a visual representation of how significant are the gains obtained are by imposing unidirectional movements and the separation of runners and walkers


\section{Exposure to droplet trails}

Recent work on aerodynamics indicates that a trail of potentially contagious droplets is left behind walking and running people, extending, respectively, for 5~m or 10~m~\cite{blocken2020towards}. That result and the present work points to the deleterious effect of running on the spread of diseases by air droplets.

In comparison to the two-way, imposing one-way movement with constant speed does not reduce the time spent on the droplet clouds, since that time is proportional to the fraction of the trail covered by the clouds, that does not depend on the direction of movement. This time can be reduced if people move faster when overpassing others, and reduce their speed when been overpassed.

If only a small fraction of the population is contagious, imposing one-way walking reduces the number of people exposed by the same proportion as the reduction in the number of crossings. However, with constant speed, the average of the total time spent in droplets clouds is not changed, i.e., in \added{one-way trails, each walker is exposed to less contagious people, but stay longer inside each cloud, with inverse proportionality.} \deleted{two-way trails, each walker is exposed to more contagious people but each contact is proportionally shorter on time.} \added{The transmission probability is reduced on one-way traffic under the assumption that it depends not only on the time spent inside the droplet cloud but increases with the number of people met, due factors such as meeting and interacting with an acquaintances, physical contact, and the variability of the host~\cite{ajelli2017host} and pathogen~\cite{sankale1995intrapatient}.}\deleted{Under the assumption that the basic reproductive rate, $R_0$,depends more strongly on the number of contacts than on the exposure time of each contact.}

\section{Conclusions}

If one-way movement and walking-only rules are imposed on bidirectional tracks shared by walkers and runners, the number of people crossing each other per minute is reduced to one-seventh of its original value and the number of crossings per 100~m is reduced to one-sixth of its original value. If one-way movement is imposed on a walking-only walkway, sidewalks, for example, the number of crossings is reduced to one-fifth of its original value. The improvements are also significant for running-only tracks. \textbf{Therefore, establishing one-way walkways and separating runners from walkers are effective measures to reduce the physical encounter of people in contagious epidemics.} 

\section*{Competing interests}

Being being myself a jogger, I am not personally pleased by the conclusions of this work.

\twoversions{\bibliographystyle{frontiersinHLTH&FPHY}}
\bibliography{pedestrians}


\end{document}